# Local excitation of kagome spin ice magnetism in HoAgGe seen by scanning tunneling microscopy

Hanbin Deng,[1,*] Tianyu Yang,[1,*] Guowei Liu,[1,*] Lu Liu,[1,*] Lingxiao Zhao,[1,2,*] Wu Wang,[1] Tiantian Li,[1,3] Wei Song,[1] Titus Neupert,[4] Xiang-Rui Liu,[1,5] Jifeng Shao,[5] Y. Y. Zhao,[2] Nan Xu,[6] Hao Deng,[7] Li Huang,[1] Yue Zhao,[1,5] Liyuan Zhang,[1,2] Jia-Wei Mei,[1,5] Liusuo Wu,[1,8] Jiaqing He,[1] Qihang Liu,[1,5,8,†] Chang Liu,[1,5] Jia-Xin Yin[1,2,†]

[1]Department of Physics, Southern University of Science and Technology, Shenzhen, China.
[2]Quantum Science Center of Guangdong-Hong Kong-Macao Greater Bay Area (Guangdong), Shenzhen, China.
[3]Shenzhen Key Laboratory of Advanced Quantum Functional Materials and Devices, Southern University of Science and Technology, Shenzhen, China.
[4]Department of Physics, University of Zurich, Winterthurerstrasse, Zurich, Switzerland.
[5]Shenzhen Institute for Quantum Science and Engineering, Southern University of Science and Technology, Shenzhen, Guangdong, China.
[6]Institute for Advanced Studies, Wuhan University, Wuhan 430072, China.
[7]School of Physical Science and Technology, ShanghaiTech University, Shanghai 201210, China.
[8]Guangdong Provincial Key Laboratory of Computational Science and Material Design, Southern University of Science and Technology, Shenzhen, China.

*These authors contributed equally to this work.
†Corresponding authors. E-mail: yinjx@sustech.edu.cn; liuqh@sustech.edu.cn

**The kagome spin ice can host frustrated magnetic excitations by flipping its local spin. Under an inelastic tunneling condition, the tip in a scanning tunneling microscope can flip the local spin, and we apply this technique to kagome metal HoAgGe with a long-range ordered spin ice ground state. Away from defects, we discover a pair of pronounced dips in the local tunneling spectrum at symmetrical bias voltages with negative intensity values, serving as a striking inelastic tunneling signal. This signal disappears above the spin ice formation temperature and has a dependence on the magnetic fields, demonstrating its intimate relation with the spin ice magnetism. We provide a two-level spin-flip model to explain the tunneling dips considering the spin ice magnetism under spin-orbit coupling. Our results uncover a local emergent excitation of spin ice magnetism in a kagome metal, suggesting that local electrical field induced spin flip climbs over a barrier caused by spin-orbital locking.**

A kagome lattice is made of corner-sharing triangles. Based on its special geometry, the kagome lattice can host frustrated magnetic or electronic structures, including quantum spin liquids [1], spin ices [2,3] and flat electronic bands [4,5]. Recent years have witnessed substantial advances in exploring the emergent physics of metallic kagome quantum materials using electronic structure methods [4,5], highlighting the interplay between geometry, topology, correlation, and spin. However, the quantum spin liquid and spin ice phase in the kagome lattice are largely unexplored by advanced electronic structure methods. In this work, we advance this research frontier by detecting pronounced tunneling conductance dips in kagome spin ice HoAgGe, serving as the signature of local spin-flip that is crucial for harnessing frustrated spin ice excitations.

In spin ice materials, the local magnetic spins respect "ice rules" similar to the electrical dipoles in water ice [2,4,6-13]. The spin ice is typically found in quantum materials with three-dimensional pyrochlore lattices, and has been recently discovered in a kagome metal HoAgGe [13]. In a kagome spin ice, the ice rule requires 2-in-1-out or 2-out-1-in spin configurations in each triangle [Fig. 1(a)]. Upon lowering the temperature, it is believed that spins first form a short-range ice order and then a long-range ice order. Intriguingly, when locally breaking the ice rule by flipping a certain spin, spin configurations with three-in or three-out can emerge, in analogy to the magnetic monopoles predicted by Dirac [24]. Creating spin excitations in a controlled and local manner can open new avenues for spin ice research.

HoAgGe consists of distorted $Ho^{3+}$ kagome lattices with layer stacking [Fig. 1(b)], where the strong local easy-axis anisotropy together with the ferromagnetic nearest-neighbor coupling of the $Ho^{3+}$ moments lead to "1-in-2-out" or "2-in-1-out" ice rules on the kagome lattice [Fig. 1(c)] [13]. The atomic structure of HoAgGe single crystal [Fig. 1(d)] is demonstrated by the high angle annular dark-field-scanning transmission electron microscopy in an aberration-corrected transmission electron microscopy, as shown in Fig. 1(e). The high-resolution image and corresponding elemental maps clearly show the Ho kagome structure in the *a-b* plane with the well-identified Ho, Ag and Ge atoms. Its short-range spin ice order occurs at around 20K, while its long-range spin ice order occurs at 11.6K [13]. In the spin ice state, magnetization versus external magnetic fields parallel to the kagome plane exhibits a series of plateaus [Fig. 1(f)], and the corresponding field-induced phases detected by neutron scattering are all consistent with the spin ice rule [13]. Magnetization along the *c*-axis, on the other hand, shows a gradual increment with field strength [Fig. 1(f)] that is consistent with a ferromagnetic canting of the spins toward the *c*-axis [13]. The temperature-dependent resistivity data [Fig. 1(g)] shows a peak feature near the spin ice transition and demonstrates metallic behavior. Consistently, our angle-resolved photoemission spectroscopy characterization of this material at the spin ice state ($T$ = 8K) reveals both Fermi surfaces and multiple dispersive bands at the Fermi level [Fig. 1(h)]. The magnetization plateaus and dispersive bands together demonstrate the coexistence of local moments and itinerant electrons. This situation allows us to use the electronic tunneling method to introduce spin excitations that we further explore.

Compared to other techniques, scanning tunneling microscope can flip the spin of a single atom through inelastic tunneling [25-31], and is highly desirable to be applied to study kagome spin ice [Fig. 2(a)]. Above a threshold tunneling bias voltage, electrons can transfer energy to the spin-flip excitations. In this work, we explore this frontier in kagome spin ice HoAgGe. Owing to the strong interlayer bonding, it is extremely rare to obtain a flat atomic surface by cryogenic cleaving, which severely challenges its scanning tunneling microscopy study. Therefore, we have to cleave more than 50 crystals and scan extensively for each cleaved crystal to find the atomic flat surface with unit-cell step heights [Figs. 2(b) and (c)]. The surface exhibits a hexagonal lattice with its lattice constant consistent with that of the crystal [Fig. 2(d)]. In the crystal structure, only Ge atoms in the HoGe layer form a hexagonal lattice. Thus, we identify this surface as HoGe surface termination. At low-temperature $T$ = 0.3K, we measure the differential conductance spectrum in the clean lattice region and detect a pair of pronounced dips located at $E_d$ = ±107meV [Fig. 2(e)], with corresponding kinks in the tunneling current data [Fig. 2(f)]. Such dips are the most pronounced features in the data and extra tiny spectral dips (not reaching to negative value) and bumps are currently not well understood. The energy locations of this pair of dips are symmetric, and their corresponding differential conductance values are strongly negative. These two features indicate the dips are from inelastic tunneling rather than from the density of states of the band structure, as also confirmed by the absence of such a feature around -0.1eV in our photoemission data in Fig. 1(h). These titanic spectral features are different from the mild step features for inelastic tunneling of spin-flip of an isolated magnetic atom [25-29]. To understand their origin, we systematically examine their spatial, temperature, and magnetic responses in Fig. 3.

We perform a spectroscopic map at the dip energy for a large area in Figs. 3(a) and (b). The spectroscopic map data is inhomogeneous, and the low-intensity parts of the map correspond to the emergence of dips, as further supported by representative differential conductance spectra along a line cut [Fig. 3(d)]. Owing to the strong interlayer bonding, the cleavage leaves vacancies and adatoms as surface defects, and we find that the dips are generically away from these defects as illustrated in Fig. 3(c), where we mark the positions of both the impurities and dips. Hence the local dip feature can be an intrinsic feature of the clean lattice rather than from impurities, and the inhomogeneity is largely caused by cleavage-induced surface defects that have a long-range suppression of the dip feature. When we raise the temperature, we find that while the dip energy does not change, the dip becomes progressively shallow, and eventually disappears above 20K [Fig. 3(e)]. It is close to the short-range spin ice order transition temperature, as evidenced by the differential heat capacity [reproduced in the inset of Fig. 3(e)]. At the defect-free region, we further find that the external magnetic field also does not change dip energy [Fig. 3(f)], but only affect its depth. An in-plane magnetization (the sample riches the 1/3 magnetization plateau with $B = 1.5T$ applied along the $b$-axis [13,32]) enhances the depth of the dip, while an out-of-plane magnetization (the sample riches the magnetization plateau with $B = 6T$ applied along the $c$-axis) suppresses the depth of the dip. We note that after applying the magnetic field or varying the temperature, we relax the system and relocate our tip to the same atomic position to take the differential spectrum. We clarify that these data are representative data rather than statistical data. The spatial, temperature, and magnetic field responses thus support the intimate relation between the pronounced tunneling dip and spin ice magnetism of the kagome lattice, and we discuss its possible microscopic origin.

In terms of magnetic excitations, the common inelastic tunneling signal can result from the spin-flip process [25-29]. The spin-flip energy $\Delta_S$ is related to the magnetic exchange interaction on the order of 1~10meV, inferred from the short-range spin ice transition temperature of 20K. As Ho carries heavy $4f$ orbitals with a large spin-orbit coupling, we need to consider the orbital flip locked to the spin-flip. Since there is no apparent orbital ordering from transport, the finial orbital flipped state would not alter the dip energy too much; however, the intermediate orthogonal orbital configuration can be at much higher energy in the anisotropic crystal field, causing an energy barrier $\Delta_O$. Therefore, the spin-flip process here can be described by a two-level system with a small spin energy difference of $\Delta_S$ and a large orbital barrier $\Delta_O$ [Fig. 4(a)]. Notably, a similar vibrationally induced two-level picture has been pointed out to explain spikes in differential conductance in molecular junctions [33]. This model further assumes different tunneling conductance values for the two energy levels, and it can naturally produce a pair of tunneling dips at symmetrical energies, as illustrated by our simulations [Fig. 4(b)], which strongly supports our two-level interpretation of the data.

To demonstrate the two-level spin-flip scenario, we present quantitative calculations of the two essential energy levels, i.e., $\Delta_S$ and $\Delta_O$. Based on the estimation [13] of local exchange interactions J from the magnetization data, we calculate the lowest spin-flip energy $\Delta_S=2J_2S^2=2$meV as shown in Fig. 4(c), where $J_2 = 0.23$meV and $S = 2$. The calculated $\Delta_S$ aligns well, in terms of the order of magnitude, with the extracted values presented in Fig. 4(e). To estimate the energy barrier of spin-flip $\Delta_O$, we calculate the crystal electric field (CEF) of $Ho^{3+}$ ion on the cleavage HoGe surface, where the surrounding ligands are $Ge^{4-}$ and $Ho^{3+}$ ions (note that it differs from the bulk CEF). The $|J = 8, J_z\rangle$ multiplet ($S = 2$, $L = 6$) of the $Ho^{3+}$ ion, under the $C_{2v}$ site symmetry of $Ho^{3+}$ ion, splits into 17 singlets [34]. We find that the ground state of $Ho^{3+}$ ion manifests Ising anisotropy, with its local axis lying in the *ab* plane, aligning with the spin ice state in HoAgGe. Moreover, we note that CEF excitations would be accompanied by spin-flip. In particular, the highest CEF level, with its magnetic easy axis perpendicular to that of the ground state, contributes to the energy barrier during the spin-flip process. This barrier is estimated to be $\Delta_O=102$ meV [Fig. 4(d)], yielding good quantitative agreement with the observed dips at $E_d=\pm 107$ meV.

In addition, as illustrated in Fig. 4(e), the two-level spin-flip model well reproduces the dependence of tunneling dips on the external magnetic field. Notably, the external field mainly changes the spin-flip energy $\Delta_S$. The 1.5T in-plane field alone would not break the in-plane ice rule (at the 1/3 magnetization plateau), but can assist spin-flip of inelastic tunneling for half of the spins in the system by introducing Zeeman energy. Thus, it reduces the minimum spin-flip energy $\Delta_S$ of the total spin system. As for the out-of-plane magnetic field, it will polarize spins to tilt towards the *c*-axis. Consequently, the $\Delta_S$ now has an additional energy contribution from the nearest magnetic exchange $J_1$, leading to a moderate increase in its value.

In summary, we extend the study of kagome materials with electronic structure methods to a kagome spin ice material, and detect pronounced tunneling dips as the signature of local spin flip. We show that the local spin-flip physics in a magnetic lattice through inelastic tunneling is different from the case of a single magnetic atom, and is well described by a two-level model involving a crystal field barrier. Our interpretation of the data leads us to conclude that effective magnetic monopoles can be pairwise created via inelastic tunneling using a scanning tunneling microscope tip. It is also meaningful to further explore the pronounced dip signal with a spin-polarized tunneling tip [35], since the spin-polarized tunneling has been achieved in other kagome materials [36-38]. Therefore, our exploration of the local spin-flip process in a spin ice system provides an initial step towards harnessing its unusual spin excitations including the magnetic monopoles. The gigantic tunneling signal in a kagome spin ice with classical spins opens up new research opportunities that encourage us to further explore the kagome spin liquid with electronic structure tools.

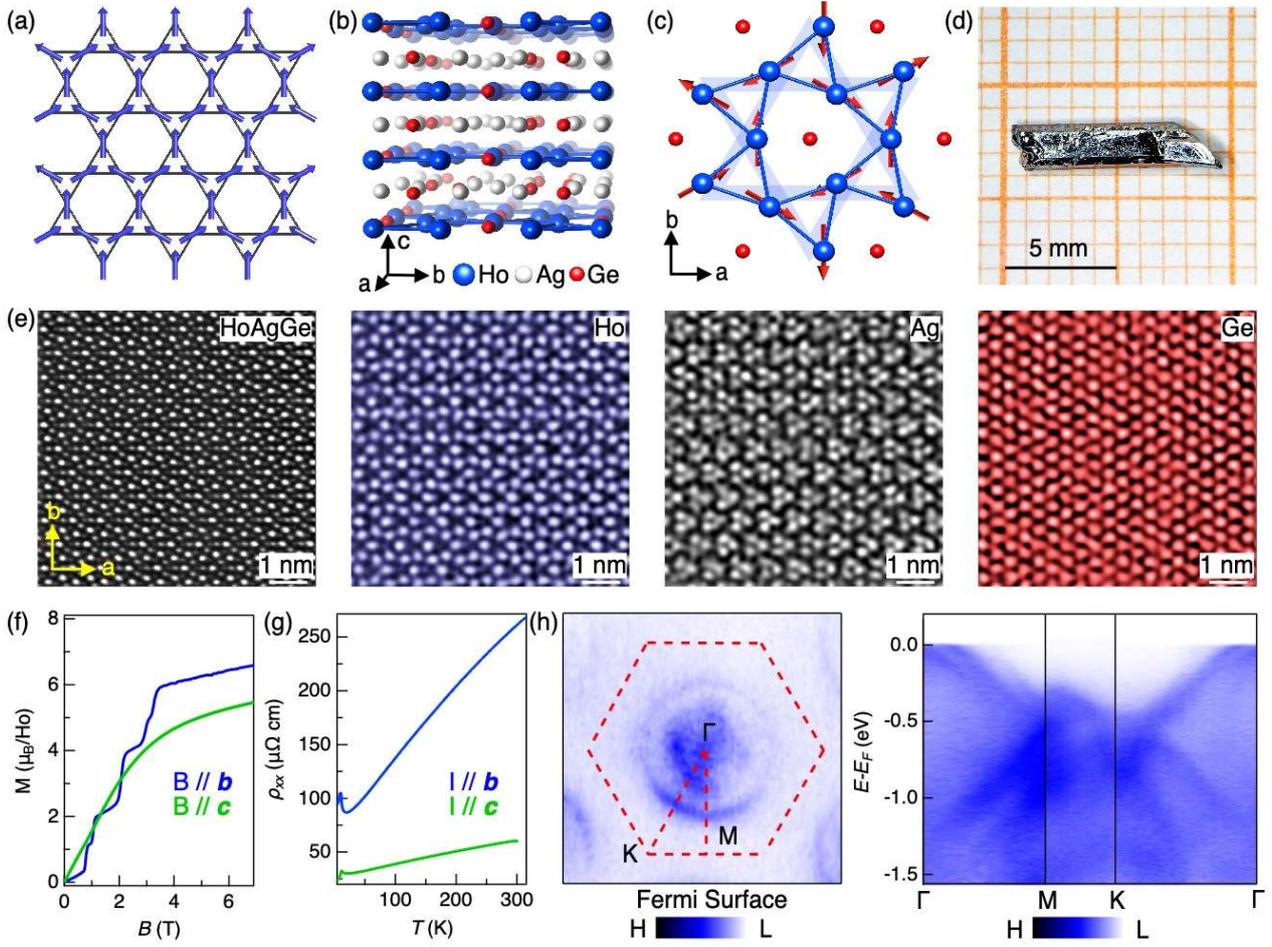

FIG. 1. Coexistence of local moment and itinerant electrons in kagome spin ice HoAgGe. (a) Schematic for a kagome spin ice. (b) Crystal structure of HoAgGe showing layered structure. (c) The distorted Ho kagome lattice with spins following the ice rule, the shaded lines show a perfect kagome lattice. (d) Image of a HoAgGe single crystal with its *c*-axis along the horizontal direction. (e) High-angle annular dark-field-scanning transmission electron microscopy image (left) and corresponding energy-dispersive spectrometry elemental maps (right), imaged along *c*-axis. (f) Low-temperature magnetization curves measured at 2K, where the in-plane magnetization curve shows fractional field-induced transitions. (g) Resistivity curve showing metallic behavior of HoAgGe. (h) Fermi surfaces (left) and band dispersions along the high symmetry lines (right) detected by angle-resolved photoemission at 8K.

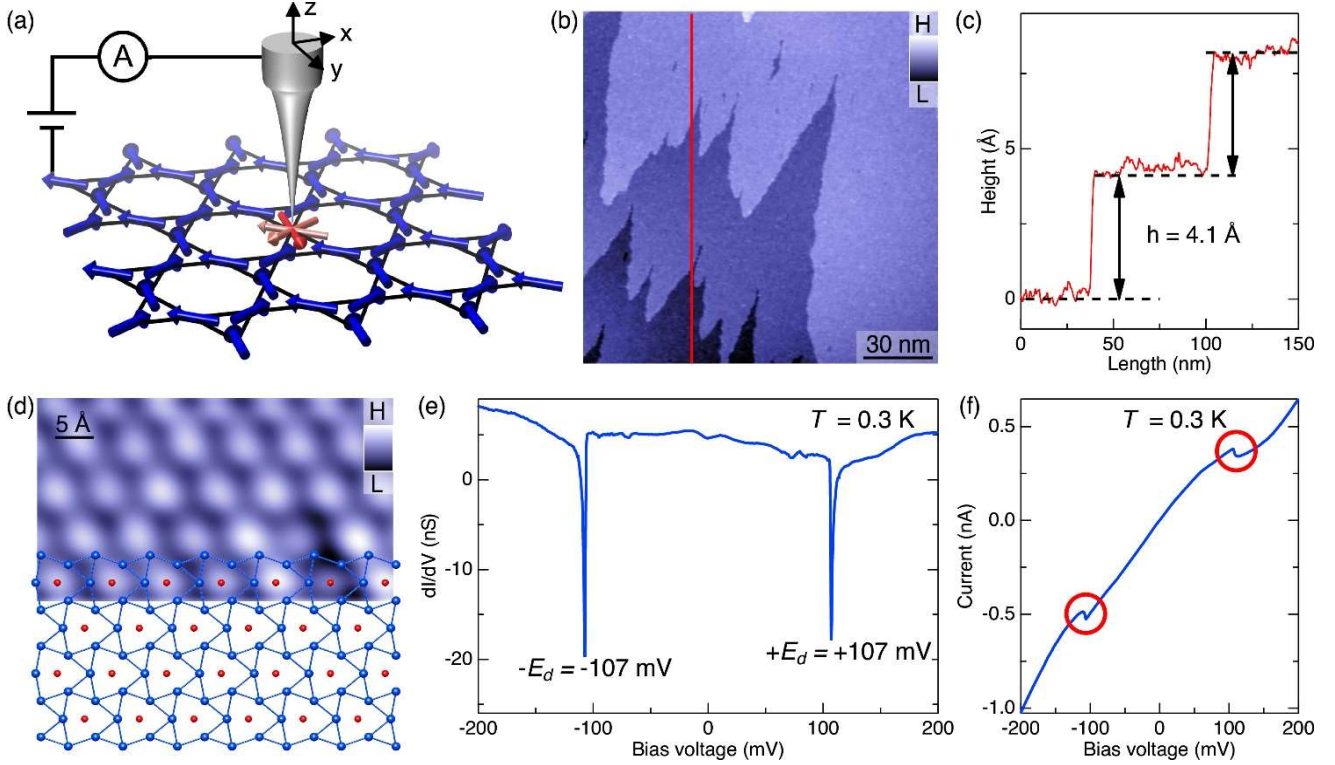

FIG. 2. Pronounced tunneling dips in kagome spin ice HoAgGe. (a) Schematic for scanning tip probe induced local spin-flip in a kagome spin ice. (b) Topographic image for a large field of view showing atomic flat cleaving surfaces with unit-cell height steps. (c) Height profile along the red line drawn in (b), showing atomic step height of 4.1Å that is consistent with the unit cell height. (d) A zoomed-in image of the topographic image (up) whose lattice symmetry is consistent with the HoGe atomic layer (down). (e) Differential conductance spectrum on the clean region of HoGe surface showing a pair of pronounced electronic dips occurring at $E_d = \pm 107$mV. (f) The corresponding kinks in the tunneling current spectrum.

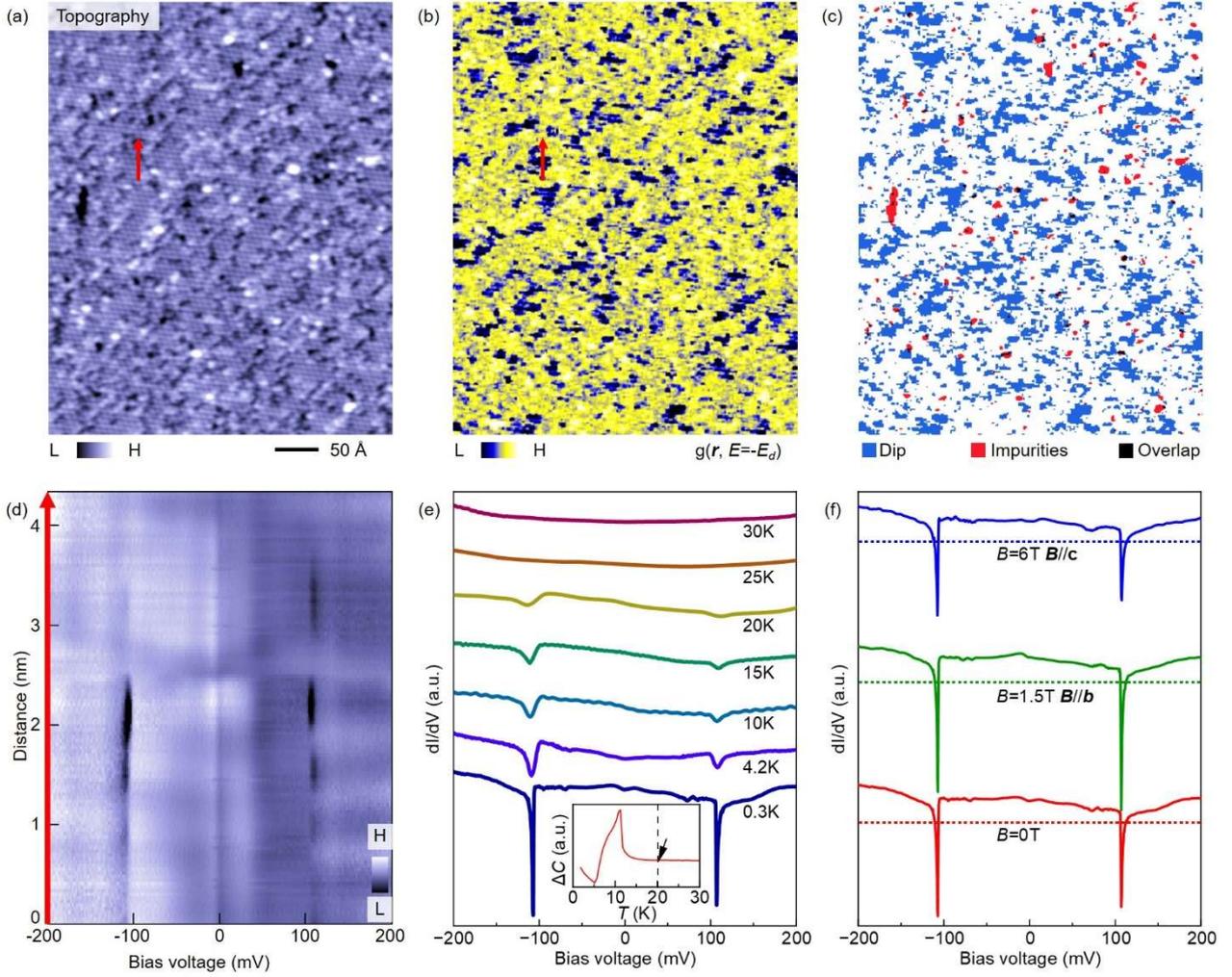

FIG. 3. Spatial, temperature, and magnetic field dependence of the tunneling dips. (a) Topographic image of a large cleavage surface. The surface contains adatoms and vacancies as impurities caused by cleavage. (b) Differential conductance map g($r$, E) at the dip energy, showing inhomogeneity. (c) Anticorrelation between the dips and impurities. We find the dips are generally away from the impurities, indicating that dips are intrinsic features of the clean lattice. (d) Intensity plot of a series of differential conductance spectra along the red line drawn in (a) and (b), showing the spatial inhomogeneity of dips at $E_d = \pm 107$ meV. (e) Temperature dependence of the dips at a fixed position, showing its disappearance above 20K. The inset shows the differential heat capacity reproduced from Ref. 13, where the short-range spin ice formation temperature is estimated to be around 20K. (f) Magnetic field dependence of the dip at a fixed position. An in-plane magnetization (the sample riches the 1/3 magnetization plateau with $B = 1.5$T applied along the $b$-axis) enhances the depth of the dip, while an out-of-plane magnetization (the sample riches the magnetization plateau with $B = 6$T applied along the $c$-axis) suppresses the depth of the dip. The dash lines mark the respective zero-value.

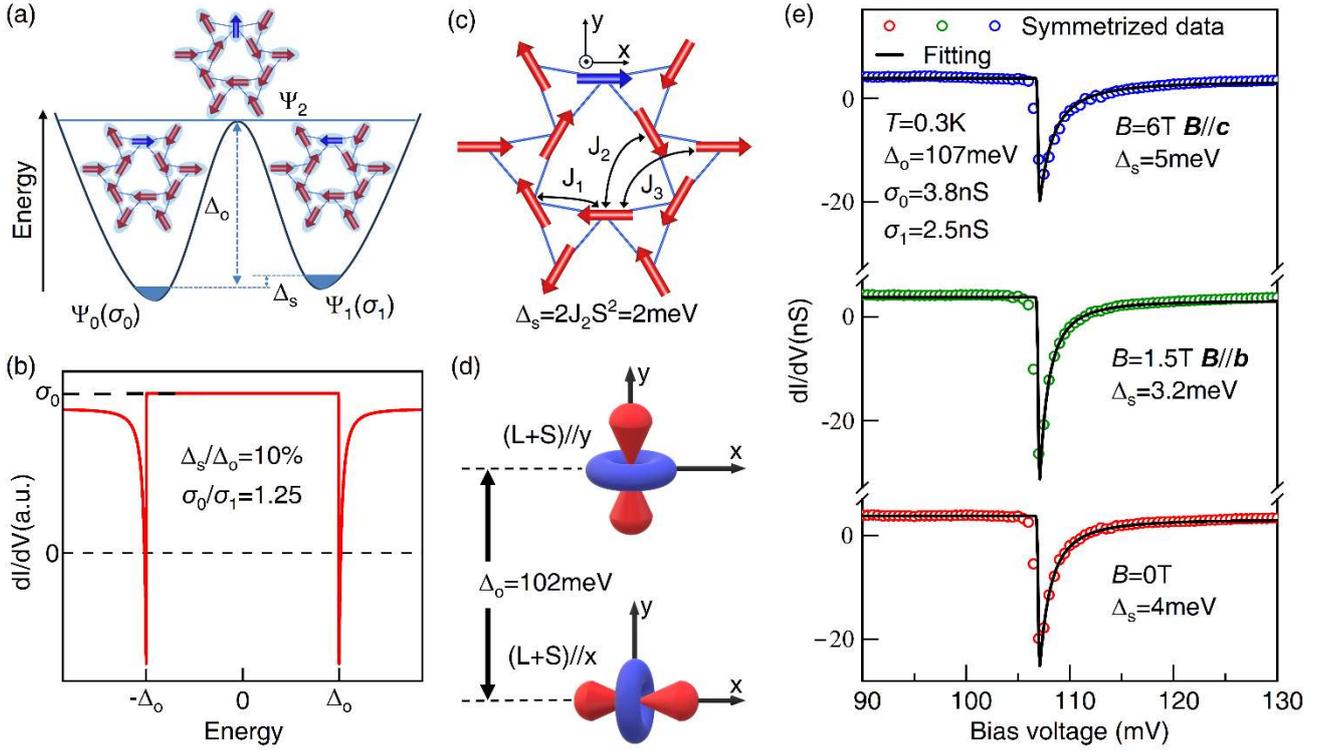

FIG. 4. Two-level spin-flip model for tunneling dips. (a) Schematic for two-level spin-flip process. In this model, $\Psi_0$ is the kagome spin ice ground state with a differential conductance $\sigma_0$, $\Psi_1$ corresponds to the spin-flipped state with a differential conductance $\sigma_1$, and $\Delta_S = 1\sim10$ meV is their spin energy difference. $\Psi_2$ is the intermediate state with orthogonal spin and orbital configurations, and it has a large energy difference $\Delta_O \sim 0.1$ eV from the anisotropic crystal electrical field. The three inset figures illustrate three different states by flipping a local spin in the distorted kagome lattice. (b) A typical simulation result of the two-level spin-flip model showing a pair of tunneling dips. (c) Illustration of three superexchange interactions $J_{1,2,3}$ in the distorted kagome lattice. The minimum spin flip energy is calculated to be 2meV based on the estimated $J$ values from the magnetization data[13]. (d) Calculated orbital barrier energy $\Delta_O = 102$ meV from the largest crystal electric field energy level. (e) Fitting results of the magnetic field dependent tunneling dips by the two-level spin-flip model. The experimental data are symmetrized with respect to zero-energy.

**Acknowledgement**
We thank the insightful discussions with Kan Zhao, Xi Dai, Yifeng Yang, and Guangming Zhang. We thank R. J. Cava for sharing his experience of using "titanic" in describing the magnetoresistance in WTe$_2$. We acknowledge the support from the National Key R&D Program of China (No. 2023YFA1407300, 2023YFF0718403, and 2022YFA1403700), National Science Foundation of China (No. 12374060, 11804402, 12004123 and 12334002), Shenzhen Fundamental Research Program (No. JCYJ20220818100405013 and JCYJ20230807093204010), the Outstanding Talents Training Fund in Shenzhen (No. 202108), Guangdong Basic and Applied Basic Research Foundation (No. 2024A1515030118), Innovative Team of General Higher Educational Institutes in Guangdong Province (No. 2020KCXTD001), Shenzhen Science and Technology Program (No. 20231117091158001), and Guangdong Provincial Quantum Science Strategic Initiative (GDZX2201001). TN acknowledges support from the Swiss National Science Foundation (Project 200021E_198011) as part of the FOR 5249 (QUAST) led by the Deutsche Forschungsgemeinschaft (DFG, German Research Foundation).

**Supplementary information**

**Methods**

<u>Single crystal growth, transport, and magnetization measurement</u>

Single crystals of HoAgGe were grown by the self-flux method. High purity elemental of Ho pieces (99.99%), Ag grain (99.999%), and Ge powder (99.999%) were weighed by stoichiometric ratio Ho:Ag:Ge = 6:705:235. The mixture was transferred into an alumina crucible and sealed in a vacuumed quartz tube. Then the tube was heated up to 1150°C for 30 h, followed by a slow cooling down to 850°C within 300 h. Single crystals of HoAgGe with hexagonal prism type were separated by centrifugation. The transport measurements were performed on a physical property measurement system (Quantum Design PPMS-14) and a variable temperature system equipped with a lock-in amplifier (NF Corp. LI5650) using the standard four-probe and alternating current transport methods. The AC current ($I_{ac}$) with frequency f = 23 Hz was applied along a-axis and c-axis respectively.

Magnetic measurements were performed in a superconducting quantum interference device magnetometer (Quantum Design MPMS-7).

Transmission electron microscopy
The thin lamella of HoAgGe single-crystal was prepared by focused ion beam along <0001> zone axis with a final 5 kV beam to minimize the sample damage. The microstructure of the prepared sample was investigated by high angle annular dark-field-scanning transmission electron microscopy (HAADF-STEM) in double Cs-corrected TEM (FEI Titan Themis G2) operated at 300 kV, with a super-X energy-dispersive spectrometry (EDS) system. Atomic-scale STEM-EDS mappings were generated in Velox (v3.90) with the radial wiener filter to enhance the feature of elemental maps.

Scanning tunneling microscopy
Single crystals with sizes up to 1mm×1mm×1mm were cleaved mechanically in situ at 80K in ultra-high vacuum conditions, and then immediately inserted into the microscope head, already at He4 base temperature (4.2K). We then further cool the microscope head to 0.3K via a He3-based single-shot refrigerator. The magnetic field was applied with a small ramping speed of 1T per 20mins. After ramping the field to a desired value, the superconducting magnet is set in the persistent mode, after which we wait for 1~2h for the system to relax and then find the same atomic position and start to take spectroscopic measurements. Tunneling conductance spectra were obtained with Ir/Pt tips using standard lock-in amplifier techniques with a root mean square oscillation voltage of $V_m$ = 0.3meV under applied bias voltage of V = 200mV and tunneling current I = 1nA. Topographic images were taken with tunneling junction set up: V = 500~200mV I = 0.05~0.5nA. The conductance maps and gap map were obtained by taking a spectrum at each location (off feedback loop) with tunneling junction set up: V = 200mV, I = 1nA, and modulation voltage $V_m$=1mV.

Angle-resolved photoemission spectroscopy (ARPES)
Clean surfaces for ARPES measurements were obtained by cleaving HoAgGe samples in situ in a vacuum greater than $5 \times 10^{-11}$ Torr. ARPES measurements were performed on the "Dreamline" beamline of the Shanghai Synchrotron Radiation Facility, with an overall energy resolution of 20 meV, angular resolution of 0.1°. The data is taken at T = 8 K, with photon energy hν = 80 eV.

1. **Two-level spin-flip model**

We have devised a theoretical model for a two-level system, exemplified by a double-well potential structure as delineated in Fig. 4(a). The system possesses two distinct spin ground states, $\Psi_0$ and $\Psi_1$, with an energy difference of $\Delta_S$ separating them. An excited state, $\Psi_2$, with a higher energy level exists within the system, and it exhibits an energy difference of $\Delta_O$ when compared to $\Psi_0$. The tunneling conductance for electrons transitioning from the different spin ground states to the excited state are denoted as $\sigma_0$ and $\sigma_1$, respectively. The tunneling current is formulated as

$$I(V) = (n_0\sigma_0 + n_1\sigma_1 + n_e \frac{(\sigma_0 + \sigma_1)}{2})V \quad (1)$$

$n_0$, $n_1$, and $n_e$ represent the electron occupation numbers for different states. Their interrelation is given

by the expression for their interrelation is Eq.(2).

$$\frac{dn_0}{dt} = -W_{e0} + W_{0e} = 0 \qquad \frac{dn_1}{dt} = -W_{e1} + W_{1e} = 0$$

$$\frac{dn_e}{dt} = -W_{0e} + W_{e0} - W_{1e} + W_{e1} = 0$$

$$n_0 + n_1 + n_e = 1 \tag{2}$$

$W$ denotes the transition rates between distinct energy eigenstates, which are quantitatively expressed via Fermi's golden rule.

$$W_{e0} = n_0 \int_{-\infty}^{+\infty} f(\xi, eV)[1 - f(\xi - \Delta_s, eV)]dE \tag{3}$$

Under the low-temperature limit, we derive the solution for Eq.(3):

$$\int_{-\infty}^{+\infty} f(\xi, eV)[1 - f(\xi + E, eV)]dE = E + \frac{1}{4}\frac{E - e|V|}{e^{(E-e|V|)/k_BT} - 1} \quad \cdots\cdots \Delta E > 0$$

$$\frac{1}{4}\frac{-E - e|V|}{e^{(-E-e|V|)/k_BT} - 1} \quad \cdots\cdots \Delta E < 0 \tag{4}$$

By solving Eq.(2), (3), and (4), we can infer that:

$$\frac{dn_0}{dt} = -W_{e0} + W_{0e} = 0 = -n_0\varepsilon_0 + n_e\varepsilon_1 = 0 \qquad \frac{dn_1}{dt} = -W_{e1} + W_{1e} = 0 = -n_0\varepsilon_2 + n_e\varepsilon_3 = 0$$

$$n_e = 1 - n_0 - n_1$$

$$\varepsilon_0 = \frac{1}{4}\frac{\Delta_s - e|V|}{e^{(\Delta_s - e|V|)/k_BT} - 1} \qquad \varepsilon_1 = \Delta_s + \frac{1}{4}\frac{\Delta_s - e|V|}{e^{(\Delta_s - e|V|)/k_BT} - 1}$$

$$\varepsilon_2 = \frac{1}{4}\frac{(\Delta_s - \Delta_o) - e|V|}{e^{[(\Delta_s - \Delta_o) - e|V|]/k_BT} - 1} \qquad \varepsilon_3 = (\Delta_s - \Delta_o) + \frac{1}{4}\frac{(\Delta_s - \Delta_o) - e|V|}{e^{[(\Delta_s - \Delta_o) - e|V|]/k_BT} - 1} \tag{5}$$

The formula for the occupation numbers is derived through the resolution of Equation 5:

$$n_0 = \frac{\varepsilon_1\varepsilon_2}{(\varepsilon_0\varepsilon_2 + \varepsilon_0\varepsilon_3 + \varepsilon_1\varepsilon_2)} \qquad n_1 = \frac{\varepsilon_0\varepsilon_3}{(\varepsilon_0\varepsilon_2 + \varepsilon_0\varepsilon_3 + \varepsilon_1\varepsilon_2)}$$

$$n_e = \frac{\varepsilon_0\varepsilon_2}{(\varepsilon_0\varepsilon_2 + \varepsilon_0\varepsilon_3 + \varepsilon_1\varepsilon_2)} \tag{6}$$

Therefore, derived from the tunneling current Eq.(1), the differential conductance is expressed as Eq.(7).

$$\frac{dI}{dV} = \frac{(\sigma_0 + \sigma_1)}{2} + \frac{(\sigma_0 - \sigma_1)}{2}(n_0 + V\frac{dn_0}{dV}) + \frac{(\sigma_1 - \sigma_0)}{2}(n_1 + V\frac{dn_1}{dV}) \tag{7}$$

The simulated differential conductance is illustrated in Fig. 4(e).

## 2. Crystal electric field (CEF) calculation

We carry out GGA calculations [39, 40] to investigate the valence states in HoAgGe. Figure S1

presents the orbitally resolved density of states (DOS). The fully occupied Ag 4d states are located at 4-6 eV below the Fermi level, suggesting the formal Ag$^+$ valence state. Similarly, the Ge 4p states, also fully occupied, lie at 0-4 eV below the Fermi level, of which the large bandwidth is mainly due to band hybridization with both Ag 4d and Ho 4f states. Consequently, the Ho$^{3+}$ valence state is deduced, collaborated by the partially occupied 4f states crossing the Fermi level.

The Ho$^{3+}$ ion, with its 4$f^{10}$ configuration, has a total spin of 2 and a total angular momentum of 6 according to Hund's rules. Thus, the multiplet with the total momentum $J = L + S = 8$ has the lowest energy, specifically the multiplet $|J = 8, J_z\rangle$ for $J_z = -8, -7, \ldots, 8$. Within the HoGe surface, the Ho$^{3+}$ ion is subject to a CEF originating from the surrounding ligand ions Ge$^{4-}$ and Ho$^{3+}$. Given the site symmetry $C_{2v}$ of the Ho$^{3+}$ ion, the $|J = 8, J_z\rangle$ multiplet would be split into 17 singlets. The CEF Hamiltonian can be written as [41],

$$H_{CEF} = \sum_{k,q} B_q^k O_q^k$$

where the $B_q^k$ are CEF parameters and $O_q^k$ are spherical tensor operators. The $C_{2v}$ site symmetry restricts the $q = 2, 4, 6$ and $k = 0, 2, 4, 6$. To estimate the CEF effect, we adopt a point-charge model via the PyCrystalField [42], setting the ligand distance range to 6 Å. The calculated CEF parameters are obtained as listed in Table 1. Moreover, the CEF eigenstates, expressed in terms of the basis $|J_z\rangle$, are tabulated in Table 2.

Note that the CEF splitting would induce magnetic anisotropy. To see this, we calculate the $3 \times 3$ matrix $M_{i,j} \equiv \langle \varphi | \hat{J}_i \hat{J}_j | \varphi \rangle$ for $i, j = x, y, z$ and $\varphi$ representing the CEF states. The magnetic easy axis of $\varphi$ is determined as the eigenvector corresponding to the largest eigenvalue of $M_{i,j}$ [43]. As a result, the ground state of the Ho$^{3+}$ ion exhibits Ising anisotropy, with its easy axis lying in the $ab$ plane. Moreover, as shown in Table 2, we find that CEF excitations may be accompanied by spin-flip. For these spin-flip states, in addition to the CEF energy, the energy contribution from the Heisenberg exchange interaction is also included. Specifically, states with an easy axis along the $y$ or $z$ direction possess an additional magnetic exchange energy of $J_2 S^2$ relative to the ground state. Using [13] $J_2 = 0.23$ meV and $S = 2$, this energy contribution is calculated to be 0.92 meV. Furthermore, our results show that the highest CEF level, energetically 101.727 meV above the ground state, features an in-plane easy axis perpendicular to that of the ground state. This aligns closely with the measured energy barrier of 107meV associated with the spin-flip process.

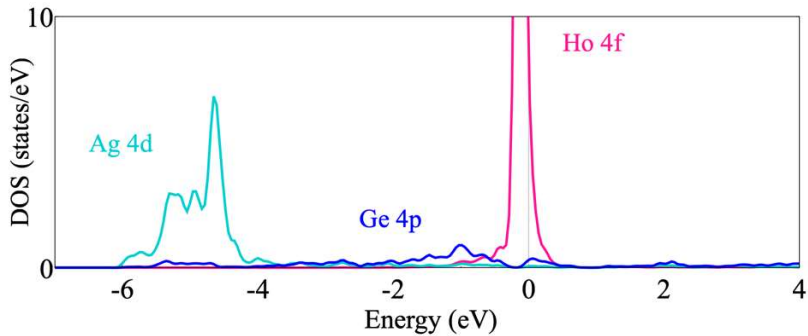

FIG. S1 The orbitally resolved DOS results of HoAgGe. The Fermi level is set at zero energy.

| $B_2^0$ | $B_2^2$ | $B_4^0$ | $B_4^2$ | $B_4^4$ | $B_6^0$ | $B_6^2$ | $B_6^4$ | $B_6^6$ |
|---|---|---|---|---|---|---|---|---|
| 1.147E-1 | -7.945E-1 | 2.970E-5 | 6.275E-4 | -9.588E-4 | 1.400E-7 | -2.240E-6 | 2.430E-6 | -4.210E-6 |

Table 1 CEF parameters $B_k^q$ (in meV) of the Ho ion within the HoGe surface of HoAgGe.

| $E$ (meV) | $\|-8\rangle$ | $\|-7\rangle$ | $\|-6\rangle$ | $\|-5\rangle$ | $\|-4\rangle$ | $\|-3\rangle$ | $\|-2\rangle$ | $\|-1\rangle$ | $\|0\rangle$ | $\|1\rangle$ | $\|2\rangle$ | $\|3\rangle$ | $\|4\rangle$ | $\|5\rangle$ | $\|6\rangle$ | $\|7\rangle$ | $\|8\rangle$ | easy axis |
|---|---|---|---|---|---|---|---|---|---|---|---|---|---|---|---|---|---|---|
| 0.000 | 0.0 | 0.0345 | 0.0 | 0.1646 | 0.0 | 0.3862 | 0.0 | 0.568 | 0.0 | 0.568 | 0.0 | 0.3862 | 0.0 | 0.1646 | 0.0 | 0.0345 | 0.0 | $x$ |
| 0.000 | 0.0096 | 0.0 | 0.0847 | 0.0 | 0.2698 | 0.0 | 0.4929 | 0.0 | 0.595 | 0.0 | 0.4929 | 0.0 | 0.2698 | 0.0 | 0.0847 | 0.0 | 0.0096 | $x$ |
| 27.365 | -0.0421 | 0.0 | -0.2534 | 0.0 | -0.4976 | 0.0 | -0.4318 | 0.0 | 0.0 | 0.0 | 0.4318 | 0.0 | 0.4976 | 0.0 | 0.2534 | 0.0 | 0.0421 | $x$ |
| 27.365 | 0.0 | 0.1261 | 0.0 | 0.3935 | 0.0 | 0.5186 | 0.0 | 0.2455 | 0.0 | -0.2455 | 0.0 | -0.5186 | 0.0 | -0.3935 | 0.0 | -0.1261 | 0.0 | $x$ |
| 45.917 | -0.1205 | 0.0 | -0.4583 | 0.0 | -0.4353 | 0.0 | 0.0782 | 0.0 | 0.3995 | 0.0 | 0.0782 | 0.0 | -0.4353 | 0.0 | -0.4583 | 0.0 | -0.1205 | $x$ |
| 45.918 | 0.0 | -0.2924 | 0.0 | -0.5246 | 0.0 | -0.207 | 0.0 | 0.3106 | 0.0 | 0.3106 | 0.0 | -0.207 | 0.0 | -0.5246 | 0.0 | -0.2924 | 0.0 | $x$ |
| 60.268 | 0.0 | 0.4864 | 0.0 | 0.3541 | 0.0 | -0.2626 | 0.0 | -0.2627 | 0.0 | 0.2627 | 0.0 | 0.2626 | 0.0 | -0.3541 | 0.0 | -0.4864 | 0.0 | $z$ |
| 60.280 | -0.2629 | 0.0 | -0.5374 | 0.0 | -0.0301 | 0.0 | 0.3757 | 0.0 | 0.0 | 0.0 | -0.3757 | 0.0 | 0.0301 | 0.0 | 0.5374 | 0.0 | 0.2629 | $z$ |
| 70.379 | 0.4782 | 0.0 | 0.3195 | 0.0 | -0.3347 | 0.0 | -0.075 | 0.0 | 0.3214 | 0.0 | -0.075 | 0.0 | -0.3347 | 0.0 | 0.3195 | 0.0 | 0.4782 | $z$ |
| 70.449 | 0.0 | 0.5861 | 0.0 | -0.0738 | 0.0 | -0.3278 | 0.0 | 0.2087 | 0.0 | 0.2087 | 0.0 | -0.3278 | 0.0 | -0.0738 | 0.0 | 0.5861 | 0.0 | $z$ |
| 76.748 | 0.6237 | 0.0 | -0.1276 | 0.0 | -0.1946 | 0.0 | 0.2383 | 0.0 | 0.0 | 0.0 | -0.2383 | 0.0 | 0.1946 | 0.0 | 0.1276 | 0.0 | -0.6237 | $z$ |
| 78.929 | 0.0 | -0.4913 | 0.0 | 0.4179 | 0.0 | -0.0638 | 0.0 | -0.2826 | 0.0 | 0.2826 | 0.0 | 0.0638 | 0.0 | -0.4179 | 0.0 | 0.4913 | 0.0 | $z$ |
| 81.109 | -0.5051 | 0.0 | 0.4026 | 0.0 | -0.184 | 0.0 | -0.1181 | 0.0 | 0.2643 | 0.0 | -0.1181 | 0.0 | -0.184 | 0.0 | 0.4026 | 0.0 | -0.5051 | $z$ |
| 88.483 | 0.0 | 0.2642 | 0.0 | -0.4385 | 0.0 | 0.4478 | 0.0 | -0.1935 | 0.0 | -0.1935 | 0.0 | 0.4478 | 0.0 | -0.4385 | 0.0 | 0.2642 | 0.0 | $y$ |
| 88.702 | -0.2001 | 0.0 | 0.3615 | 0.0 | -0.4622 | 0.0 | 0.34 | 0.0 | 0.0 | 0.0 | -0.34 | 0.0 | 0.4622 | 0.0 | -0.3615 | 0.0 | 0.2001 | $y$ |
| 101.722 | 0.0 | -0.0779 | 0.0 | 0.2124 | 0.0 | -0.3975 | 0.0 | 0.5393 | 0.0 | -0.5393 | 0.0 | 0.3975 | 0.0 | -0.2124 | 0.0 | 0.0779 | 0.0 | $y$ |
| 101.727 | 0.0398 | 0.0 | -0.1364 | 0.0 | 0.303 | 0.0 | -0.481 | 0.0 | 0.5596 | 0.0 | -0.481 | 0.0 | 0.303 | 0.0 | -0.1364 | 0.0 | 0.0398 | $y$ |

Table 2 The energy, wave functions and magnetic easy axis of 17 CEF singlets.

Extended STM data analysis

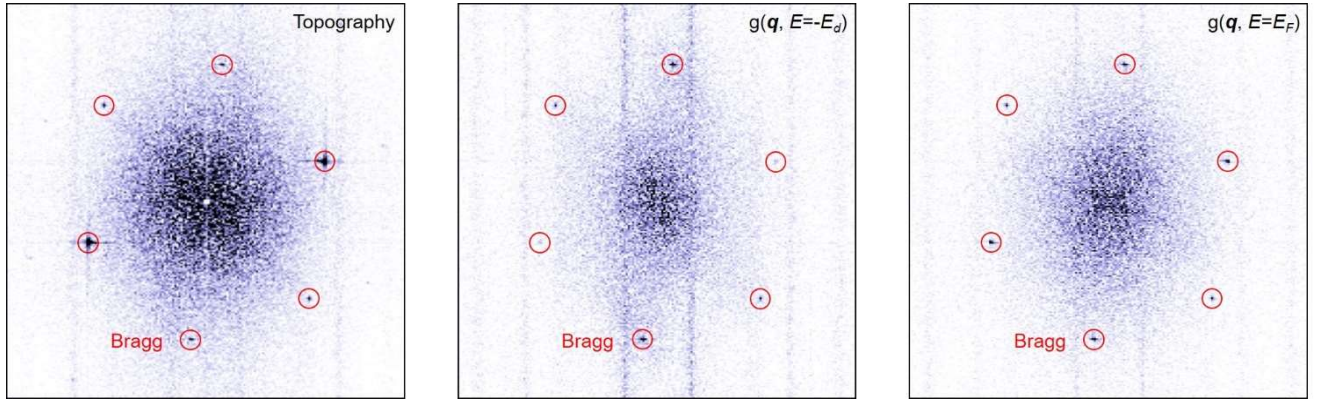

FIG. S2 Fourier transform analysis. We show the Fourier transform analysis for the map data set in the area of Fig. 3, which include the Fourier transform of topographic data (left), map data at the dip energy (middle) and map data at zero-energy (right). In all three cases, the clearest signal is the Bragg peaks marked by red circles, indicating the major spatial modulation of the electronic states comes from the lattice. The vertical stripes are from the instrumental noise. Add data are taken at 300mK.

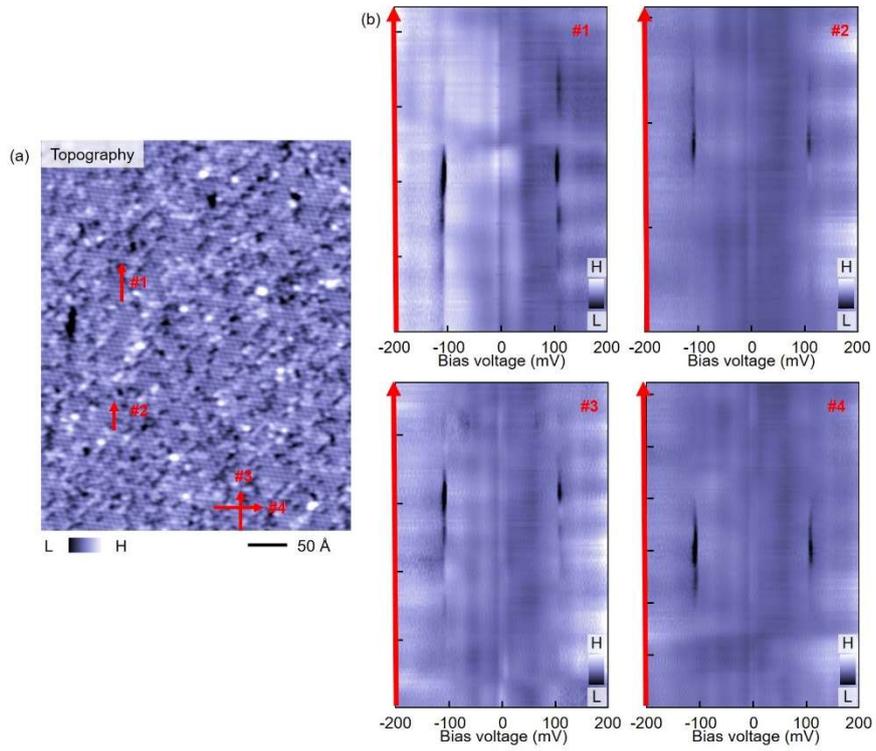

FIG. S3 Additional linecut spectrums (b) taken on the position marked on the topographic image in (a). Add data are taken at 300mK.